# SUMMARY OF THE NIJMEGEN WORKSHOP ON MULTIPARTICLE PRODUCTION

RUDOLPH C. HWA

*Institute of Theoretical Science and Department of Physics*
*University of Oregon, Eugene, OR 97403-5203, USA*
*E-mail: hwa@oregon.uoregon.edu*

## 1 Introduction

This has been a highly successful workshop in which many new results were presented and there was ample time for extended discussions. A notable feature of this workshop is that, despite its small size, there are many participants from a wide variety of countries, many representing the younger generation. With the data of 56 participants from 22 countries I made for amusement a single-event intermittency analysis of the participation distribution using a maximum of 55 bins corresponding to the number of countries listed in our data-diary blue book. I found an approximate power-law behavior with $\varphi_2 = 0.24$ and $\varphi_3 = 0.62$. It would be interesting to see how these intermittency indices will vary in future workshops in this series.

At the planning stage of this workshop I expressed to W. Kittel my worry about no new experiments on fluctuations, while he responded with his worry that there were no new developments in theory. As it turns out, I believe that we were both somewhat too pessimistic. Let me organize this summary in the following way. I partition the topics into three categories: (a) phenomenology of mature topics, (b) experiments not driven by conventional theory, and (c) theory not driven by conventional experiments. With improved resolution the subtopics are: ($a_1$) Bose-Einstein correlations, ($a_2$) fluctuations, ($a_3$) phenomenology of QCD and other dynamics, ($b_1$) DCC, ($b_2$) soft photons, ($c_1$) critical behavior, and ($c_2$) chaos.

## 2 Phenomenology of Mature Topics

### 2.1 Bose-Einstein Correlations

The subject of Bose-Einstein correlations (BEC) has been around for a long time, at least thirty years just in particle physics alone. In recent years the number of parameters used to describe the sources of multiparticle production has increased rapidly, indicating a growth and maturing period. According to Weiner there is still much work to be done, as he outlined a program that will extend the investigation well into the next century.[1]

One of the hindrances to the development of this subject is that there are many conventions and notations used by many people with diverse backgrounds, writing a large number of papers that do not attempt to relate to one another. Weiner [1] tells us that there are in general 10 independent parameters in $C_2$, but it is not easy to see how they are related to, for example, the "standard form" for the parameterization by CSH.[2] Perhaps it is a good sign that there is not yet a standard model so that many flowers can bloom, but it would be nice if some standard notation can be established.

Apart from the details, the significance of the recent development is clear. HBT interferometry offers a way to probe the time evolution of the emitting sources that no other method can provide.[3] Depending on the geometrical symmetry and the dynamical properties of the expanding system, various forms of the correlation function $C_2$ can be derived. Of course, the more elaborate the theoretical input, the more parameters will appear in $C_2$, and therefore it is more likely that the data can be fitted. A case in point is the work of EHS/NA22 as reported by Hakobyan.[4] It is found that the BEC data of hadronic collisions at $\sqrt{s} = 22$ GeV do not support a volume-emitting fireball-like source of spherically symmetric Gaussian space-time distribution; nor are they fitted very well by the Bowler parameterization for a string-line source. What works best is the model based on a hydrodynamically expanding cylindrical source with specific inputs on the time dependences of the longitudinal and transverse source radii.[2,5] While the result is interesting in that it reveals the nonstatic nature of a nonspherical source, it seems hard to accept the accuracy of the detailed dynamical inputs such as Bjorken's boost-invariant flow and freeze-out temperature $T_f = 140$ MeV, applied to hadronic collisions where the cm energy is so low that $\langle n_{ch} \rangle$ is only about 8, $dn/dy$ not very flat, and the relevance of the hydrodynamics of a thermal system is questionable.

For heavy-ion collisions one would have more confidence in the applicability of the hydrodynamical description. Seyboth [6] reported on the results of the NA49 experiment on Pb-Pb collisions. By assuming that unlike-sign correlation $C_2^{+-}$ is due entirely to the Coulomb interaction, they obtain a good fit with a rather large radius of $R = 4.6$ fm. Then after applying that Coulomb correction to the like-sign correlation $C_2^{--}$, they get (a) $R_{\text{long}}$ to be independent of $y$ in the longitudinal cm system, (b) $R_{\text{side}}$ and $R_{\text{T}}$ around 6 fm (roughly the nuclear radius of Pb), (c) freeze-out time $\tau_0$ around 6.5 fm/$c$, and (d) duration of decoupling $\Delta\tau$ around 2.5 fm/$c$. Because $\Delta\tau < \tau_0$, it is concluded that there is no evidence for a first-order phase transition. The data also indicate that the genuine multiparticle ($\geq 3$) correlations are weak.

On the matter of genuine 3-particle correlations (which cannot be expressed in terms of the 2-particle correlations, i.e., the irreducible component), DELPHI has found them in the hadrons produced in $e^+e^-$ annihilation.[7] It is of interest to note that the radius parameter for the genuine portion of $C_3$ is $> 3$ fm, while that for $C_2$ is $< 1$ fm. The proper understanding of that fact has not been suggested by anyone so far. Why does the higher-order correlation take place when the system is larger? It has also been found in the DELPHI data that $R_{t,\text{out}}$, $R_{t,\text{side}}$ and $R_{\text{long}}$ all decrease with increasing transverse mass as $A + B/\sqrt{m_t}$.[8] While that may be reasonable in nuclear collisions due to the higher density gradient at smaller $R$ and earlier time, why should it be also the case for $e^+e^-$ annihilation?

Andersson gave a colorful review of the basics of the Lund model for the purpose of describing how the matrix elements can be symmetrized to yield the effects of BEC in the model.[9] Results of the upgraded model have not yet been compared to the correlation data. Bańas gave an elegant review of the theoretical foundation of correlations, relating particle spectrum to density matrix and then to Wigner functions.[10] He then proposed a way to modify an algorithm for the probability distribution in order to generate events with identical particles by use of an approximation of the Wigner function.

Picking up on the experimental support of the longitudinally expanding source by the NA22 data,[4] Csörgő discussed the phenomenological consequences of expanding vs nonexpanding sources. Then he described a halo model that takes into account long-lived resonances.[11] In my view there are two complementary aspects of the BEC problem. One is to understand the details of the time evolution of a large, hydrodynamical system by the interferometry method, of which Csörgő's work is an example. The other is to use the correlation functions to discover new properties of the system whose dynamics is unknown, or at least unconventional. The latter is exemplified by the talk of Eggers,[12] who uses the $\bar{p}p$ data of UA1 to illustrate his point. The two seem to manifest themselves in nonoverlapping regions of the momenta-difference squared $Q^2$ so that the former at larger $Q^2$ has proceeded independently of the latter at lower $Q^2$. Eggers reported on the study of the cumulant data of UA1 in $-\log Q^2$, confirming the power-law behavior at small $Q^2$. Then in attempting to check the theoretical constraints of quantum statistics, it is found that the second and third order cumulants of the UA1 data cannot be simultaneously fitted by simple parameterizations (Gaussian, exponential and power) of the normalized field correlators $d_{ij}$ in the theory. The result suggests the inadequacy of the conventional description of BEC in terms of a combination of chaotic and coherent components consisting of only 2-particle correlators.

## 2.2 *Fluctuations*

The subject of fluctuations has undergone a swing of attention that appears to be oscillatory. Recall that initially intermittency referred to the scaling behavior of the normalized factorial moments $F_q$ in the resolution size $\delta$.[13] Then it was shown that the statistics of the same effect can be dramatically improved by the consideration of the correlation integrals,[14] which are plotted as functions of $-\ln Q^2$. Once the variable $Q^2$ is used, the connection with BEC is too close to be ignored. Indeed, experimentally it was shown that in like-sign pairs BEC dominates the intermittency phenomena.[15] Incidentally, that finding has been a source of great rejoicing to those who work with the dual parton model, which has had some difficulty in explaining the power-law behavior of $F_q$,[16] but was relieved of the burden when BEC was found to be the primary cause of the effect. The Fritiof model has also been amended by taking the BEC into account.[17] However, the curious, and therefore interesting, part of the physics is not removed by merely symmetrizing the states of identical particles. That is what I want to emphasize in the next few paragraphs.

First, let me recall that intermittency refers to the fluctuations of the rare, spiky events. Since $F_q$ is zero unless $n \geq q$ in a bin under consideration, only the tail of the multiplicity distribution $P_n$ can make any contribution to $F_q$, when $\delta$ is very small, i.e., when $\langle n \rangle_\delta \ll q$. Thus, a power-law behavior of $F_q$ vs $\delta$ at small $\delta$ is a statement about $P_n$ at $n \gg \langle n \rangle_\delta$. For example, for hadronic collisions where $\langle n \rangle_\delta \sim 0.01$ at the smallest bin, $F_2$ picks out very rare events with at least two particles in such small bins. (That is not the case with nuclear collisions, which explains why interesting effects in intermittency are not seen at low $q$ there.) Tannenbaum's fit of $P_n$ over the whole range of $n$ by NBD is undoubtedly inaccurate at the extreme tail,[18] and therefore cannot claim to account for intermittency. This same situation is now repeated in the study of BEC.

When correlation functions are plotted against $Q^2$, they are usually well fitted by some formulae over a wide range of $Q^2$, thereby determining some BEC parameters. This is illustrated in Fig. 1 which shows the NA22 data,[19] and the fits are in accordance with a superposition of chaotic and coherent components.[20] Notice, however, that there are points at the smallest $Q^2$ values that cannot be fitted. If the same data are exhibited in the log-log plot, as shown by Kittel[21] in Fig. 2, then the BEC fits are clearly inadequate at large $-\ln Q^2$. In fact, the data points appear to favor straight-line behavior without saturation, highly reminiscent of intermittency. Thus the conventional BEC does not account for the whole story. The unconventional, power-law behaved part can still be an aspect of BEC, as Białas would insist, but it is the unconven-tional aspect that contains new physics and should not be overlooked in the same breath that dismisses intermittency on the basis of Fig. 1.

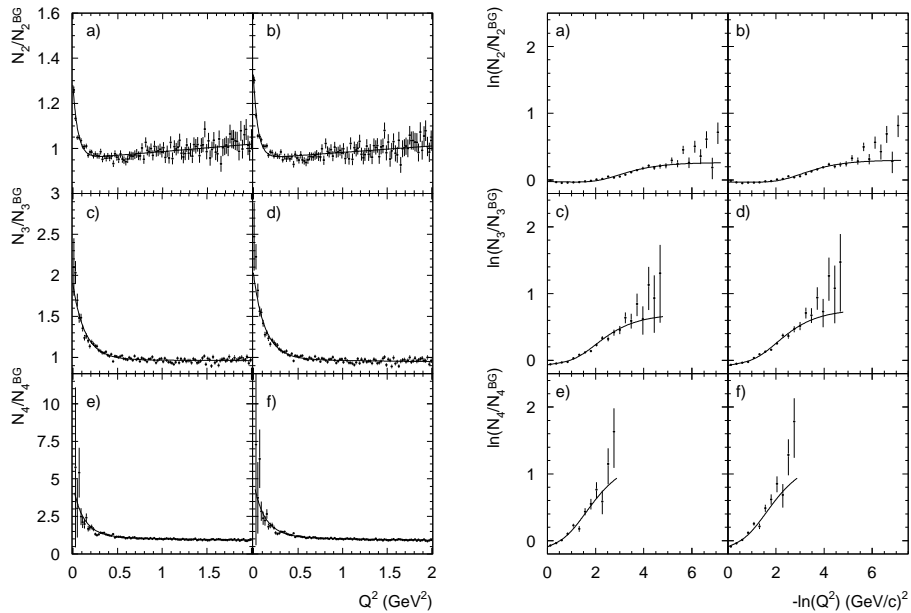

Figure 1: Data of NA22 for the normalized 2-, 3- and 4-particle inclusive densities not corrected (a,c,e) and corrected (b,d,f) for Coulomb interaction. Curves show the fits by linear combinations of Gaussian distributions. See Ref. 19.

Figure 2: Same as in Fig. 1 but plotted as functions of $-\ln Q^2$.

BEC refers to like-sign charged particles. What about the unlike-sign ones? The log-log plots of the correlation integrals $F_q^S$ vs $Q^2$ are shown in Fig. 3 for different sign combinations.[22] Clearly, the unlike-charge data for $F_2^S$ exhibit significant power-law behavior, though weaker than the like-charge data. Their difference may well correspond to the rising solid lines in Fig. 2 due to the conventional BEC effects. If so, then the unconventional part is the same in both the like and unlike charge sectors, revealing the self-similar nature of its dynamical origin. Of course, precise statements on the subject cannot be made until the effects of resonance contributions, $\gamma$ conversions, etc., are clearly understood and taken into account. For another example on the same point, we show the third-order cumulant $K_3$ in Fig. 4 to indicate the power-law behavior of various charge combinations in the NA22 data.[23]

Having presented the experimental evidence for interesting fluctuations beyond the conventional BEC, let us now see what has been discussed on the measures of fluctuations. Blažek, Seixas, Czyżewski, Wu and Liu have all commented on the properties of $F_q$ in one way or another.[24] Space limitation does not permit comments on them all; they represent refinements of the more familiar measures. On the intermittency indices $\varphi_q$, Ziaja discussed the relationship between the average of $\varphi_q$ and the theoretical $\varphi_q$ in the $\alpha$ model.[25] Seyboth reported on the weakness of the intermittency signal in the NA49 Pb-Pb data,[6] while Grinbaum-Sarkisyan found more structure in the low $A$ (C-Cu collisions) and low energy (4.5 $A$ GeV/$c$) data.[26]

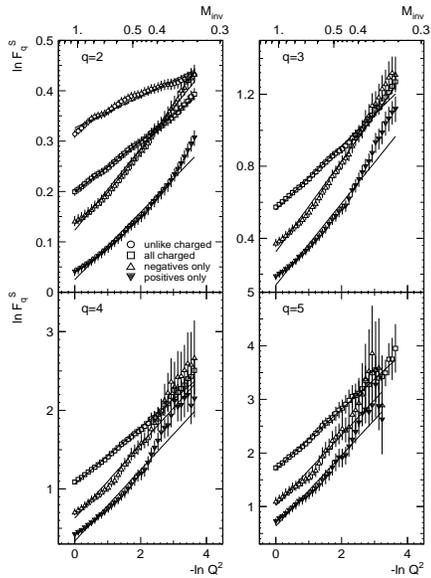
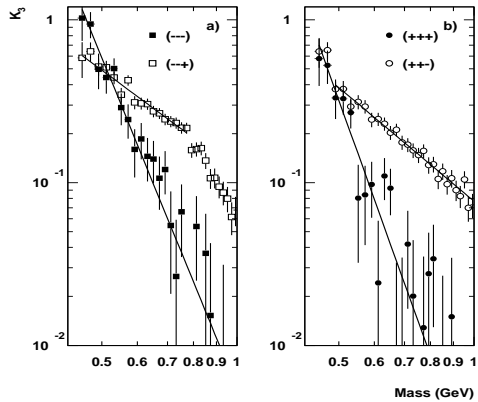

Figure 3: Data of NA22 for the correlation integrals for various charge combinations. See Ref. 22.

Figure 4: Data of NA22 for the third-order normalized factorial cumulants for various charge combinations. See Ref. 23.

The virtues of wavelet analysis have been extolled by its proponents for a number of years now.[27] Greiner gave a very lucid description of it again this time.[28] Conceptually, it is reasonable to expect the successive dilating and shifting of the multiresolution analysis to have definite advantages over the Fourier analysis, say, when applied to multiparticle production. However, in practice its superiority has not been demonstrated in the analysis of real data. Without such demonstration there are few converts who are willing to invest the necessary effort to learn the intricacies of the new "mathematical microscope". Sarcevic[29] and Biyajima[30] both have applied the wavelet analysis to the JACEE events, and got very different results. It may be because they have studied different events. But it is not clear what one is supposed to have learned from the results of their analyses, especially when they are so drastically different. Is there any universal feature that has been uncovered, as Białas and Peschanski[13] did in treating the same data?

A very interesting new development this year is the discovery of the bunching parameters.[31] Their definition is

$$\eta_q(\delta) = \frac{q}{q-1}\frac{P_q(\delta)P_{q-2}(\delta)}{P_{q-1}^2(\delta)} \quad . \tag{1}$$

Like $F_q$ they are 1 for Poissonian distribution, but are more sensitive to some details of the fluctuations. In the limit $\delta \to 0$ they are related to $F_q$ by

$$F_q(\delta) \simeq \prod_{i=2}^{q} \eta_i^{q-i+1}(\delta), \qquad \delta \to 0 \quad . \tag{2}$$

Their superiority over $F_q$ is exemplified by Figs. 5 and 6, where $\eta_q$ show dissimilar behaviors in the multiplicity fluctuations generated by the JETSET 7.4 PS model, when analyzed in bins of the azimuthal angles defined with respect to the beam axis (Fig. 5a) vs those to the thrust axis (Fig. 5b).[32] However, $F_q$ in Figs. 6a and b show very similar behaviors for the same two cases. Not only are the appropriate angles identified by Fig. 5b, but also the power-law behaviors of $\eta_q$ are more in evidence. Although the bunching parameters cannot be directly related to the generating functions (which bridge between theory and experiment), they seem to have sufficient merits to deserve further study.

### 2.3 Phenomenology of QCD and Other Dynamics

Meunier[33] described the connection between pQCD and the $\alpha$ model, which has become the standard toy model of fragmentation that gives rise to intermittency.[13] The $\alpha$ model itself is related to the random energy model in statistical physics through the common use of the Cayley tree.[34] Since these connections illuminate the basis of the evolution equation in pQCD, it is worth going through briefly here the various key links involved.

A Cayley tree is a generic branching diagram, which for simplicity we take to have only two branches at each vertex. In statistical physics such a tree is taken seriously as a geometrical structure on which one considers self-avoiding walks, i.e., walking from the top (root) to the bottom (without retracting) and making choices at each vertex. In the random energy model[34] there is a random potential $V$ with a probability distribution $\rho(V)$ that influences the choice of path on the tree. Without going into the details of that problem that involves partition function, inverse temperature, etc., let us accept an intuitive formula on the generating function $G_j(u)$, where $j$ represents the $j$th step of the walk on the tree,

$$G_{j+1}(u) = \int dV \rho(V) \left[G_j(u+V)\right]^2 \quad . \tag{3}$$

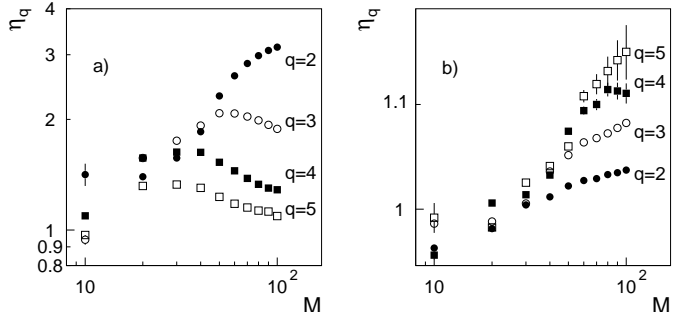

Figure 5: Bunching parameters as functions of the number of bins in the azimuthal angle $\phi$ defined with respect to the a) beam axis and b) thrust axis (JETSET 7.4 PS). See Ref. 32.

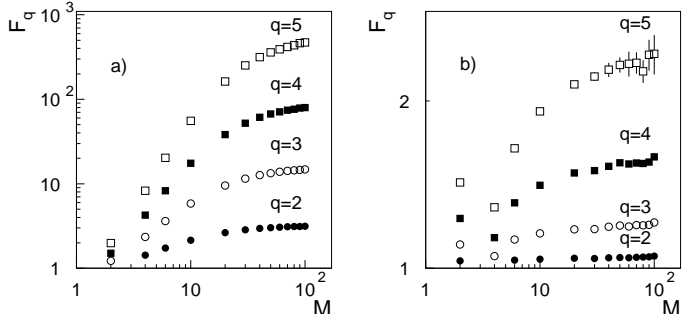

Figure 6: Normalized factorial moments plotted as in Fig. 5. See Ref. 32.

This equation can be given the interpretation that the generating function for a $(j+1)$-step path is related to those for two $j$-step paths, joined together at the first vertex (at the root), integrated over the distribution on $V$ that influences the decision made there. If we change the discrete variable $j$ to a continuous variable $t$, and change $G_{j+1}(u) - G_j(u)$ to $\partial H(t,u)/\partial t$, then we can see a similarity between Eq. 3 and

$$\frac{\partial}{\partial t} H(t,u) = \int dW \, r(W) H(t, uW) H(t, uW'(W)) - H(t,u) \quad , \tag{4}$$

where $W$ is another sampling variable. Eq. 4 is the evolution equation of the $\alpha$ model,[33,35] where $r(W)$ is the normalized weight probability distribution for branching, and $W'(W)$ reflects some constraint on the density splitting.

To relate Eq. 4 to the QCD evolution equation

$$\frac{\partial G(Q,u)}{\partial \ln Q} = \int \mathrm{d}x K(x) \left[ G(Qx,u)G(Q(1-x),u) - G(Q,u) \right] \qquad (5)$$

requires a key link that can connect $H(t, uW)$ to $G(Qx, u)$. The main point is to see how the integration variable can be moved from the second to the first variable, i.e., $G(Q, xu) \to G(\lambda Q, u)$. That is accomplished by the ansatz [33] $G(Q, u) = g(u \langle n \rangle_Q)$, which is consistent with KNO scaling,[36] as will be shown below. Because $g(u \langle n \rangle_Q)$ is a function of only one variable, we therefore have $G(\lambda Q, u) = g(u \langle n \rangle_{\lambda Q}) = g(xu \langle n \rangle_Q) = G(Q, xu)$, where $x = \langle n \rangle_{\lambda Q} / \langle n \rangle_Q$. Hence, without going into the details of the factors involved, we can see that Eqs. 4 and 5 are very close, provided that the above ansatz is acceptable. To see that, recall that the second factorial moment is

$$f_2 = \langle n(n-1) \rangle_Q = \left. \frac{\mathrm{d}^2}{\mathrm{d}u^2} G(Q, u) \right|_{u=1} = \langle n \rangle_Q^2 \, g'', \qquad (6)$$

which is a statement of KNO scaling if $g''$ is nearly independent of energy.

On the subject of fragmentation models Zborovský[37] described the properties of the multiplicity distributions by solving the rate equations that include the birth and death terms of partons. Such an approach was tried many years ago by a number of people.[38] It is not clear what advances have been made by Zborovský and how the deficiencies encountered by the earlier work are overcome, e.g., the treatment of infrared and collinear divergences in QCD. Płoszajczak,[39] on the other hand, seems to have dealt with the problem and solved the fragmentation-inactivation kinetic rate equations that contain the QCD equations in MLLA.

On angular intermittency one works with angles $\theta, \delta$ and replaces $Q$ in Eq. 5 by $P\theta$.[40] In fact, instead of $\theta$, a variable $\varepsilon = \ln(\theta/\delta)/\ln(P\theta/\Lambda)$ is found to be more natural. Phenomenology in $\varepsilon$ has been reported in this workshop (see below).

Dremin[41] has in recent years been emphasizing the advantages of studying the cumulant moments $K_q$, or better still the ratio $H_q = K_q/F_q$. The reasons are that Eq. 5 can be solved only under certain approximations and that the result, when expressed in terms of $H_q$, is highly sensitive to the approximation made. In particular, $H_q$ oscillates in $q$, a theoretical prediction that has been seen in the experiment.[42,43]

The origin of the observed oscillations turns out to involve other factors beside the dynamics of QCD. It is known that for negative binomial distribution $H_q$ decreases monotonically with increasing $q$.[44] However, if the generating function $G(z)$ is defined as a finite sum over $P_n^{NB}$, i.e., $G(z) = \sum_{n=0}^{N}(1+z)^n P_n^{NB}$, where $N$ is finite, then the corresponding $F_q$ and $K_q$ derived from such a $G(z)$ by $q$th order derivatives will yield oscillating $H_q$.[45] Furthermore, a linear superposition of two $P_n^{NB}$ that can better fit a shoulder [46] in $P_n$ will give rise to oscillating $H_q$ also.[47] Ugoccioni,[48] speaking in place of Giovannini, described how the data on $H_q$ from SLD that show the oscillations [42] can be fitted by an appropriate combination of both truncation and superposition. Thus the original impact of the data on oscillations, regarded as a support for the NLLA on QCD calculations, has been weakened.

The problem about the zeros of the generating function continues to be fascinating and puzzling. When Lee and Yang tried to understand the mathematics of phase transition, they found that when the sum in the grand partition function, $Q_N(z) = \sum_{n=0}^{N} z^n Z_n$, for a lattice gas is truncated at a large, but finite $N$, and under certain conditions for the interaction between two gas molecules, then the $N$ zeros of $Q_N(z)$ lie on the unit circle in the complex $z$ plane.[49] De Wolf was the first one to notice that when the $P_n$ from JETSET-PS is used in the generating function, $G_N(z) = \sum_{n=0}^{N}(1+z)^n P_n$ its zeros also form a nearly perfect circle.[50] Dremin and Gianini [51] now show that the $\bar{p}p$ data of UA5 also exhibit the Lee-Yang circle. De Wolf has questioned what the effect of the errors in the real data would be on the circle. To me it is not clear why they are circles in the first place for multiparticle production, nor what we can learn from the result. That is not to say that they are not interesting. On the contrary, any mystery like that is worth further study.

Ochs[52] discussed the soft limit of the energy spectrum of gluon emission in QCD, and gave the not-so-surprising analytical result that there exists a scaling limit. His earlier work with Wosiek [40] on the angular correlation in QCD jets has a prediction on the correlation ratio $r(\varepsilon)$ as a function of the variable $\varepsilon$ mentioned earlier. That dependence on $\varepsilon$ has been found by Buschbeck [53] to fit well the DELPHI data by the choice of $\Lambda = 0.3$ GeV.

## 3 Experiments not Driven by Conventional Theory

### 3.1 Search for DCC

The disoriented chiral condensate (DCC) is certainly not a conventional idea.[54] It is conjectured that under certain conditions of a high-energy collision a region of vacuum can be created in which the chiral orientation is different from that of the normal vacuum outside, such as in the Baked Alaska scenario.[55] When the DCC meets the normal vacuum with the standard chiral orientation, the pions radiated can have any orientation in the isospin space, i.e., $dn/d\cos\theta$ is a constant, where $\theta$ is the polar angle in isospace. Since the fraction of neutral pions is $f = \cos^2\theta$, one gets $n^{-1}dn/df = 1/(2\sqrt{f})$. That is the spectacular signature of the DCC: a preponderance of charged pions with small $f$. Andreev, however, warns us that the $1/\sqrt{f}$ behavior is only a necessary consequence of DCC, but not sufficient; a wide class of coherent and squeezed states can give rise to such an $f$ distribution also.[56,57] Whatever the origin may be, the observation of any novel charge fluctuations in the laboratory would be highly stimulating.

Like C. Columbus, Bjorken sets out to search for DCC by building his own ship.[58] Certain parameters ($r_2, r_3$, etc.) that are ratios of appropriate normalized double factorial moments, involving charged and photon multiplicities, are calculated from the data of their Fermilab experiment (T864), and used as signatures of DCC. For example, $r_2$ for the conventional mechanism of multi-particle production is 1, but for the unconventional DCC mechanism it should be 1/2. The data of MiniMax so far yields $r_2 = 0.95 \pm 0.01$. Too bad, but more analysis needs to be done. The preliminary conclusion is that DCC has not been found.

Since the MiniMax detector is limited to a very small part of the phase space, it may not be looking at the place where DCC is most likely to be produced. But given the detector, one may question whether the parameters such as $r_2$ involve too much averaging. If DCC is produced, but only very rarely, then one should make data analysis that is sensitive to large fluctuations from event to event. A plot of the $f$ distribution may be more revealing, since a little bump at small $f$ may stand out above the generic background, and yet makes negligible contribution to the overall average.

Sarcevic[29] described a study that shows how the DCC signal can be identified by the use of wavelet analysis.[28] In the study the linear $\sigma$-model is used to generate the DCC events with various initial conditions.[59] The DCC signal can be distinguished from the background because the discrete wavelet transform exhibits different behaviors for them at different scales. The method is more effective at high rapidity density, and is therefore more suitable for heavy-ion collisions than hadronic collisions like the ones studied by MiniMax. My general feeling about the theoretical work done thus far on the DCC is that far more light has been shed on its evolution and decay characteristics than on its production mechanism. Some crucial conjecture has been made concerning the origin of the chiral disorientation. Unless we know better how the DCC is created and protected, if only briefly, in a realistic high-energy collision, one cannot maximize the chances of detecting its existence. In that sense the project is far more difficult than proving that the earth is round.

## 3.2 Soft-Photon Production

Spyropoulou-Strassinaki has been reporting over the past several years about the excess of soft-photon production over the expected rate from radiative hadronic decays and hadronic bremsstrahlung.[60] The excess has been seen in WA27, NA22, EMC and WA83; now she has preliminary results from new experiments (WA83 and WA91*) confirming the previous results. The photon energy is in the range $0.1 - 1$ GeV and $p_T < 80$ MeV/$c$. Using the Fritiof Monte Carlo to simulate the photons from hadronic decays, they find that the difference between the data and Fritiof simulation is about 10 times greater than the expected hadronic bremsstrahlung rate, although the dependences on $p_T$ are rather similar. There has been no adequate theoretical explanation of this phenomenon. The success in discovering what is unexpected has been sufficient to drive the experiments. But where are the theorists?

## 4 Theory not Driven by Conventional Experiments

### 4.1 Criticality

The question of whether intermittency is a signature of critical behavior was raised again, this time by Peschanski.[61] On the basis of model studies of nuclear fragmentation his answer is a tentative yes, i.e., there is some sign of criticality. Presumably he was only looking for suggestions, since hints are all what can be expected from specific fragmentation models. It is known that there are other models where criticality does not give rise to conventional intermittency.[62] The meaning of criticality was also asked — in the context of multiparticle production. The necessity for a tunable dimensionless parameter, implied by the answer, seems to rule out a whole class of phenomena, called self-organized criticality,[63] the relevance of which to quark-hadron phase transition has already been found possible.[64] Nevertheless, the meaning of criticality for pQCD processes is definitely worth questioning and investigating.

Antoniou described his work on observable fluctuations at phase transition, which is the latest of a long series of studies on chiral condensates.[65] My difficulties in appreciating the work are connected to my lack of understanding of how the observables enter into the theoretical formalism. At $T = T_c$ there are, strictly speaking, no condensates except by fluctuations, so it is not clear what the observable normalized factorial moments are, when $\langle n \rangle_\delta$ is a nontrivial quantity that cannot be calculated from an effective action. Furthermore, at freeze-out, $\tau_f \gg \tau_c$, all the signatures from phase transition may be smeared out by the thermal system in the hadron phase between $\tau_c$ and $\tau_f$.

In the Ginzburg-Landau description of phase transition where the order parameter is related to the hadron density,[62] there is no scaling behavior in $\delta$, but $F$-scaling is valid for a range of $T < T_c$, i.e., $F_q \propto F_2^{\beta_q}$, where $\beta_q = (q-1)^\nu$ with $\nu \simeq 1.3$. Efforts to introduce spatial fluctuations in order to generalize the result beyond mean-field theory have not led to any significant alterations in the value of $\nu$.[66]

### 4.2 Erraticity and Chaos

In primitive times we measured $P_n$ and determined $\langle n \rangle$ and $\langle n^2 \rangle / \langle n \rangle^2$. In more modern times we studied the dependence of $F_q$ in $\delta$, where $F_q$ involves both horizontal and vertical averages. Since such averages destroy any information on the spatial pattern of an event, one can consider a horizontal $F_q$ for each event and study the distribution $P(F_q)$ of such horizontal $F_q$ after sampling many events. The normalized moments $C_{p,q} = \langle F_q^p \rangle / \langle F_q \rangle^p$ that involve vertical averages over $P(F_q)$ then retain some information on the fluctuations of spatial patterns.[67] Erraticity [68] refers to the scaling behavior of $C_{p,q} \propto M^{\psi_q(p)}$. Such behaviors have been found in the Monte-Carlo simulation of QCD parton showers and in the data of hadronic collisions.[51] It is a challenge for models of soft interaction to reproduce the experimental $\psi_q(p)$ or the erraticity spectrum $e_q(\alpha)$.

The issue of chaos in multiparticle production processes relies on the acceptance of a sensible measure of chaos, when the difficulty of tracking the time evolution of a self-interacting quantum system destroys the utility of the Lyapunov exponent that has been found so useful in classical nonlinear dynamics. The entropy indices $\mu_q = \mathrm{d}\psi_q(p)/\mathrm{d}p|_{p=1}$ turn out to be highly effective in providing such a measure.[67,68] Experimental data on $\mu_q$ for leptonic, hadronic and nuclear collisions are needed to help establish a deeper understanding of the nature of fluctuations. It should also be recognized that the quantification of fluctuations of spatial patterns has wide applications far beyond the realm of high energy physics — such as in biology, chemistry, and even geology.

## 5 Other Comments

Limitations in space (here) and time (in Nijmegen) force me to leave out comments on many other interesting contributions to this workshop. This statement is not meant to hide my deficiency in understanding and appreciating certain papers and my loss of concentration at a number of talks. Thus the absence of any references to those contributions in this summary is a reflection of my limitation, not theirs.

There are some general remarks I would like to make that are not related to any specific problems presented here. One concerns the Pomeron. One reason for studying multiparticle production is to learn more about soft interaction. A thorough knowledge of the production processes should by unitarity tell us more about diffractive scattering, and therefore the Pomeron. But how is that (even just in principle, if not in practice) related to the Pomeron structure, a subject that has become increasingly fashionable recently? People who are interested in the structure of the Pomeron in hard diffractive processes or in the properties of the Pomeron in small-x physics seem to be very disinterested in fluctuations of the kind discussed here, which provide clues to the nature of multiparticle production. This tendency to fission into narrower subcommunities needs to be reversed for the benefit of the whole. One possible problem that may bring them together would be to study the size of the Pomeron by BE correlation analysis of identical particles produced in double diffractive processes. Another problem that may contribute to cross-cultural linking is to investigate the possible connection between the self-similarity of Hagedorn's statistical bootstrap model [69] and that of intermittency.

A point that I wish to express is about the future of the subject of soft interaction. Whereas fluctuation is a legitimate topic of study in statistical physics, it is not viewed with enthusiasm in the general high-energy physics community. Declaring complex systems as being not fundamental does not reduce the intrinsic value of understanding complexity as a subject in its own right, especially when such complexity cannot be avoided in most high-energy experiments. There are some who feel that the subject should be connected with others in the main stream, such as jet physics, small-x physics, or heavy-ion physics, in order to gain acceptance. Indeed, such connections should be established, wherever possible, for good physics reasons. However, the pursuit of understanding soft physics should not be abandoned merely on the ground of its complexity. To me physics is more interesting when there is no "standard model". It is especially exciting when the problems require the use of a wide variety of techniques and theories developed over the whole spectrum of subfields in physics. The dynamics of quarks and gluons that we must deal with is that of a many-body system, for which we need the knowledge acquired in quantum optics, condensed-matter, statistical, nuclear and particle physics. Developments in recent years have led some to go beyond the traditional domain of particle physics and study critical phenomena, self-organized criticality, multifractality, coherent and squeezed states, spin-glass systems, and chaotic behaviors, to name a few. Soft physics viewed as the study of many-body, strongly-interacting microscopic systems is in the unique position of being able to explore the universality of physics, which is what makes the subject exciting and challenging.

Finally, let me remark that the excellent atmosphere conducive to the discussion of these problems at length would not have been possible without the tireless work of the local organizers. Most notable among them is, of course, Wolfram Kittel, under whose able leadership not only was the pleasant experience at Ons Erf made memorable, but more significantly the NA22 collaboration has become the beacon of light in soft physics. His persistent devotion to high-quality work in an unsympathetic environment makes him the Medici of multiparticle production, with the difference that he does the creative work himself.

**Acknowledgments**

I have benefitted from discussions with many participants too numerous to list here. In concrete measure my task of preparing this summary has been made significantly easier by the help of W. Kittel and the scientific secretary, S. Chekanov. This work was supported, in part, by the U. S. Department of Energy under Grant No. DE-FG03-96ER40972.